\documentclass[aps, reprint, superscriptaddress, nofootinbib]{revtex4-1}
\pdfoutput=1
\usepackage{amsmath, latexsym, amssymb, hyperref, graphicx, color, slashed,xcolor}
\definecolor{nicered}{rgb}{0.7,0.1,0.1}
\definecolor{nicegreen}{rgb}{0.1,0.5,0.1}
\hypersetup{colorlinks, citecolor=nicegreen,linkcolor=nicered}
\begin{document}

\title{Parity and the origin of neutrino mass}

\author{Goran Senjanovi\'{c} }
 \affiliation{International Centre for Theoretical Physics, Trieste, Italy}
\affiliation{Tsung-Dao Lee Institute \& Department of Physics and Astronomy, SKLPPC, Shanghai Jiao Tong
University, 800 Dongchuan Rd., Minhang, Shanghai 200240, China}

\author{ Vladimir Tello }
 \affiliation{International Centre for Theoretical Physics, Trieste, Italy}

\date{\today}

\begin{abstract}
        
  In the LHC era the issue of the origin and nature of neutrino mass has attained a new meaning and a renewed importance. The growing success of the Higgs-Weinberg mechanism behind the charged fermion masses paves the way for answering the question of neutrino mass. We have shown recently how the spontaneous breaking of parity in the context of the minimal Left-Right Symmetric Model allows to probe the origin of neutrino mass in complete analogy with the charged fermions masses in the Standard Model. We revisit here this issue and fill in the gaps left in our previous work. In particular we discuss a number of different mathematical approaches to the problem of disentangling the seesaw mechanism and show how a unique analytical solution emerges. Most important, we give all the possible expressions for the neutrino Dirac mass matrix for general values of light and heavy neutrino mass matrices. In practical terms what is achieved is an untangling of the seesaw mechanism with clear and precise predictions testable at hadron colliders such as LHC.

    \end{abstract}

\maketitle

%
%
\section{Parity and the origin of charged fermion masses} 

 There is a growing evidence that elementary particles owe their masses to the Higgs mechanism. It is now certain that this is true for the $W$ and $Z$ bosons, and the third generation of charged fermions. The way to verify it is both simple and deep: the knowledge of particle masses determines uniquely the Higgs boson decay rates. In this sense the Standard Model (SM) is a completely self-contained theory whose predictions are purely structural and do not require any additional assumptions. It is becoming safe to assume then that the Higgs mechanism works for quarks and charged leptons, and the issue becomes whether the same is true for neutrino. The origin of neutrino mass is thus arguably a great priority and our best bet for the physics Beyond the SM. After all, the vanishing of neutrino mass is the only real failure of the SM.

      \vspace{0.1cm}
    {\bf The origin of charged fermion masses.} In the case of the charged fermions, the Higgs origin of their masses is verified through the decays of the Higgs boson into fermion and anti-fermion pairs. The crucial point is that the mass fixes uniquely the Yukawa coupling 
\begin{equation}\label{SMy}
  y_f = \frac{g}{2} \frac{m_f}{M_W},                                               
\end{equation}   
which then gives the relevant decay rate
   \begin{equation}\label{Higgsdecay}
   \Gamma (h \to \bar f f ) \propto m_h (m_f/M_W)^2.                                               
\end{equation}
 In other words, the structure of the SM allows us, without any further assumptions, to associate a well defined prediction of a relevant physical process to the mass in question. Masses become dynamical parameters, and here lies the beauty of the SM. 
 
Now, where does \eqref{SMy} come from? The answer lies in the maximal parity violation which makes  left-handed (LH) fermions doublets and the right-handed (RH) fermions singlets of the $SU(2)_L$ gauge group. This forces the Higgs multiplet to be a doublet and, remarkably enough, a single doublet suffices for the masses of all SM particles except for neutrino. Hence the above connection of Yukawa couplings with the $W$-boson and fermion masses. This also explains why in the SM fermion masses stay at the electro-weak scale, instead of escaping to larger scales. Moreover, it guarantees the absence of flavor violation in neutral currents since mass and Yukawa coupling matrices are diagonalised simultaneously. To  appreciate  this, one can imagine for a moment a parity conserving world of vector-like fermions and try to construct the analog Higgs-Weinberg theory of their masses, consistent with phenomenology.  One then sees immediately that this program would fail  to account for each of the points discussed above.

The remarkable predictivity and simplicity of the SM Higgs sector is a miracle of parity violation. In the words of Weinberg: "V-A was the key"~\cite{Weinberg:2009zz}.
 Ironically, the same parity violation then leads to a vanishing neutrino mass in the SM due to the absence of the RH neutrino. In a parity conserving world, though, neutrino would be massive due to the $SU(2)$ symmetry between electron and neutrino.
 
  \vspace{0.1cm}
     {\bf The origin of neutrino mass.}
We have then a kind of catch twenty-two situation: we need maximal parity violation for the sake of charged fermion masses and at the same time parity conservation for the sake of neutrino. It is not surprising that the solution lies in the idea to break parity spontaneously, as in the left-right symmetric theory that predicted neutrino mass from the outset, long before experiment. The question is whether this theory can provide the quantitive answer to the issue of the origin of neutrino mass, the way the SM does it for charged fermions. The answer is yes, as we discuss in the following.
 
 In the SM the origin of charged fermion masses is related to the Higgs decays into fermion pairs. Understanding the origin of neutrino mass should then simply mean the same: have a theory with structural predictions for relevant particle decays associated with neutrino masses, without any additional input. 
 
    The dominant scenario today 
    behind the smallness of neutrino mass is based on the seesaw mechanism~\cite{Minkowski,Mohapatra:1979ia,rest}. Its main feature is the addition of a new heavy Majorana neutral lepton $N$ per generation (traditionally called right-handed neutrino)  to the SM. Through the  Dirac mass term $M_D$ between $\nu$ and $N$  one obtains a non-vanishing neutrino mass matrix
\begin{equation}\label{seesaw-I}
M_{\nu}=-M_D^T\frac{1}{M_N}M_D,
 \end{equation} 
 which holds true for $M_N \gg M_D$, a natural assumption for the mass matrix $M_N$ of gauge singlets $N$.

At first glance the seesaw mechanism seems to explain neutrino lightness, but that is somewhat misleading since a priori one has no idea what $M_D$ and $M_N$ are.  The way it is written it would seem that $M_\nu$ is a consequence, but in reality it is $M_\nu$ that it is being probed today in low energy experiments and  thus should clearly be an input. If $N$ states are physical, accessible to colliders, then $M_N$ can in principle be determined and used as an input. If however one imagines $N$ as unreachable ghost states that give us $M_\nu$ when integrated out, then the seesaw mechanism does little good - rather, it trades a physical question of $M_\nu$ to the unphysical one of $M_N$. In this case, one is better off with an effective d=5 operator~\cite{Weinberg:1979sa} in our opinion. 

In other words, the task is to determine $M_D$ as a function of $M_\nu$ and $M_N$, the latter in principle measurable at the LHC or a future hadron collider. The knowledge of $M_D$ would give us the $\nu-N$ mixing and in turn allows us to make predictions for the relevant $N$ decays. In what follows this is what we will imply by disentangling the seesaw mechanism or by probing the origin of neutrino mass. This is analogous to probing the origin of charged neutrino masses. 

%

It is well known that the  seesaw formula \eqref{seesaw-I} cannot be untangled since $M_D$ is determined up to an arbitrary complex orthogonal matrix. 
This obstacle should not come up as a surprise. After all, the seesaw mechanism is normally portrayed as an ad-hoc extension of the SM - the addition of gauge singlets which have arbitrary masses and couplings, unrelated to any new physical principle.  In particular, 
 it is worth to contrast it with the left-right symmetric extension \cite{PatiSalam}  of the SM that attributes the left-handed nature of weak interactions to the spontaneous breakdown of parity. It is precisely the LR symmetric theory that led originally to the existence of RH neutrinos and to the non-vanishing neutrino mass long before experiment.

In the modern version of the theory, hereafter denoted Minimal Left-Right Symmetric Model (MLRSM)~\cite{Minkowski,Mohapatra:1979ia,MohSenj81}, the seesaw mechanism follows naturally from spontaneous symmetry breaking, with $N$ mass proportional to the mass of the right-handed charged gauge boson, 
$M_N\propto M_{W_R}$. The smallness of neutrino mass is thus linked to the near maximality of parity violation in weak interactions \cite{Mohapatra:1979ia} - in the limit of infinite $M_{W_R}$ one recovers the massless neutrino of the SM. 

The crucial role in this is clearly played by the spontaneously broken left-right symmetry which has a priori two options, i.e. generalized parity $\mathcal{P}$ as we argued above, but also generalized charge conjugation $\mathcal{C}$.  
      The case of $\mathcal{C}$ is rather transparent with since it implies symmetric Dirac mass matrices of quarks and leptons. This  forces the LH and RH mixing angles in the quark sector to be the same, which is of great importance in determining the limits on the RH gauge boson mass. On the other hand, the condition $M_D^T=M_D$ in the neutrino sector, allows to determine $M_D$ as a function of $M_\nu$ and $M_N$, and thus verify the Higgs source of neutrino mass \cite{Nemevsek:2012iq}.

\vspace{0.1cm}
{\bf Restored parity: probing neutrino mass.}\label{pand nu} The case of  $\mathcal{P}$ is however more involved. The Hermitian Dirac Yukawa couplings do not imply Hermitian mass matrices due to the complex vacuum expectation values in general. In such a case, one needs an alternative approach based on the decays of doubly charged scalars and  the heavy SM doublet, since they probe directly $M_D$ \cite{Senjanovic:2016vxw}. 
  
  It turns out though that in the case of unbroken parity in the Dirac Yukawa sector, $M_D$ can be determined and for equal left and right-handed leptonic mixing matrices one gets~\cite{Senjanovic:2016vxw}
  \begin{equation}\label {MDirac}
M_D=i\, V_L \sqrt{m_{\nu}m_N}V_L^{\dagger},
\end{equation}
where $V_L$ is the PMNS mixing matrix. We chose $V_L = V_R$  for illustrative purposes since it leads to no loss of generality. The  point is that there is no ambiguity left unlike in the SM seesaw scenario discussed above.
One then gets the flavor conserving decays rates of heavy RH neutrinos
  \begin{equation}\label {Ndecay}
  \Gamma (N_i \to h \nu_i) \propto \Gamma (N_i \to Z \nu_i) \propto m_{N_i}^2 m_{\nu_i}/M_{W}^2
\end{equation}
and their flavor violating decay rates
 \begin{equation}\label {Ndecay}
\Gamma (N_i \to W^{\pm} \ell_j^{\mp}) \propto    m_{N_i}^2 m_{\nu_i}|(V_L)_{ij}|^2/M_{W}^2.
\end{equation}
The last decays are particularly striking since they probe directly the Majorana nature of $N$. Namely, once produced on-shell, the RH Majorana neutrinos $N$ must decay equally into leptons and anti-leptons~\cite{Keung:1983uu}. There are other ways of probing the Majorana nature of heavy neutral leptons when this particular channel is not available, see e.g~\cite{Balantekin:2018ukw}, but none this clear and dramatic. It should be added that the Majorana nature of $N$ may be negligible if two such almost degenerate $N's$ conspire to make a Dirac or pseudo-Dirac particle, see e.g.~\cite{Gluza:2016qqv}
 or when one uses the CP phases to achieve partial cancellations of decays into charged anti-leptons for three or more $N's$, see e.g.~\cite{Das:2017hmg}. 
.

The above expressions  illustrate the predictivity of the theory, in complete analogy with \eqref{Higgsdecay} for charged fermions and allow the hadron collider study of $M_D$; for a recent work see ~\cite{Helo:2018rll}.
  In a detailed appraisal~\cite{Senjanovic:2018xtu}, we described  the theoretical and phenomenological aspects  of parity as the LR symmetry, but nonetheless we are doubly motivated to further elaborate on our logic and our results. To start with, our solution for $M_D$ in the Hermitian limit appears somewhat mysterious and it may not be clear to the reader what made us choose the particular approach. Firstly, we wish to show here its uniqueness since other approaches are indirect and lack a clear physical picture. More important, we complete the program of finding all possible analytic solutions for $M_D$ in terms of the light and heavy neutrino masses and mixings, which finally justifies the claim of  untangling the seesaw with verifiable structural predictions.

Before we plunge into the details of our work, a comment on a phenomenological motivation for a possibly accessible LR breaking scale. 
 Imagine that the neutrinoless double beta decay is seen in near future. What would that imply? One logical and simple answer is that it is driven by the neutrino Majorana mass, but  this is far from being the only possibility. In the  case of normal neutrino mass hierarchy this is unlikely, so it the normal hierarchy was to be established meanwhile, it could as well be new physics causing this process. If electrons emitted were to be right-handed, the neutrinoless double beta decay could only come from new physics. It should be stressed that the argument for possible new physics was brought up already sixty years ago \cite{maurice}, and after all, neutrino Majorana mass implies new BSM physics. 

The neutrinoless double beta decay is a dimension nine six-fermion operator with the coefficient that scales as the fifth power of the scale of new physics (for the sake of illustration we assume a single such scale)
  \begin{equation}\label {0nu2beta}
\frac{1}{\Lambda^5} \, n \,n\, \bar p\, \bar p\, \bar e\, \bar e 
\end{equation}
From the experimental bound $\tau_{0 \nu 2 \beta} \gtrsim 10^{25}$\,yr, one can obtain the lower limit 
$\Lambda \gtrsim  3$\,TeV (it is easy to see that it corresponds to a limit $m_\nu \lesssim 1 $\,eV). Compare this with proton decay d=6 effective operator $ q q q \ell$ which due to proton longevity takes the associated scale of new physics above $10^{15}$\,GeV, completely out of direct reach.

The consequences of this are important. If neutrinoless double beta is observed and if it is induced by new physics, it could lie tantalisingly close to the LHC energies. The neutrinoless double beta decay could be a probe of the theory behind neutrino mass, and not the probe of neutrino Majorana mass itself, as often claimed. 

 For this reason, it becomes necessary to address the predictions of the MLRSM for the lepton number violation at the hadron colliders. The golden channel is the form of same sign charged lepton pairs and jets, the so-called KS process~\cite{Keung:1983uu},  with a plethora of other associated processes that could help untangle the seesaw~\cite{Senjanovic:2018xtu}.   These processes, especially the KS, serve an essential role of finding $M_N$ which then together with $M_\nu$ serves as an input for predicting $M_D$ and associated decays. In this sense, $M_\nu$ and $M_N$ are the analog of $m_f$ for a charged fermion. Neutrinos, being Majorana particles provide a more complex system with both light and heavy states, but the principle and the resulting physics are basically the same.

  The central part of this note is given in section \ref{section:crux} where we show how to find all solutions for $M_D$ without any additional assumption whatsoever. First however, in the next section we summarize the main features of the MLRSM so that the reader can follow the technical aspects of the section \ref{section:crux}. Our conclusions  are left for the section \ref{section:outlook} where we also comment on alternative approaches to the issue of neutrino mass. 
  
  To have a brief picture of what probing the origin of neutrino mass is all about, see~\cite{Senjanovic:2016pza}.

\section{Minimal Left-Right Symmetric Model with Parity}

In this section we give the most salient features of the theory needed in order to present our results regarding neutrino mass, the relevant details can be found in~\cite{Senjanovic:2018xtu}. A reader in need of a more in-depth review of the theory would benefit from~\cite{Tello:2012qda}.

The MLRSM is based on the following symmetry group 
\begin{equation}\label{LRgroup}
              \mathcal{G}_{LR} = SU(2)_L \times SU(2)_R \times U(1)_{B-L}  \times P                                                  
              \end{equation}
where a discrete generalized parity  $\mathcal{P}$ plays the role of left-right symmetry.
We need discuss only the leptonic sector (the quark sector is discussed in~\cite{Senjanovic:2014pva}). 
The LH and RH lepton doublets
\begin{equation}
  \ell_{L,R} = \left( \begin{array}{c} \nu \\ e \end{array}\right)_{L,R}.                              
         \end{equation}
transform under $\mathcal{P}$ as
\begin{equation}\label{ellunderP}
             \ell_L \leftrightarrow \ell_R.                                    
\end{equation}

In order  to break the original symmetry down to the SM one at the large scale, one needs left and right $SU(2)$ triplets
  $ \Delta_L (3,1,2)$ and $\Delta_R (1,3,2)$,                
where the numbers in brackets denote the representation content under \eqref{LRgroup}. These scalars have the following form    
\begin{equation}
\Delta_{L,R}=\left(
 \begin{array}{c c}
  \delta_{L,R}^+ /\sqrt{2}& \delta_{L,R}^{++} \\ [3pt]
\delta_{L,R}^0 & - \delta_{L,R}^+ /\sqrt{2}
\end{array} 
\right)
\end{equation}
Prior to the SM gauge symmetry breaking, one has $v_L=\langle \delta_L^0\rangle= 0, v_R=\langle \delta_R^0\rangle\neq 0$. The vev $v_R$  gives masses to the heavy gauge bosons $W_R$ and $Z_R$ and the RH neutrinos $N$, leading to  the breaking of $G_{LR}$ down to the SM gauge symmetry.

The Standard Model gauge symmetry breaking is achieved by the $SU(2)_L \times SU(2)_R$ bi-doublet $\Phi (2,2,0)$, containing two SM doublets 
\begin{equation}
\Phi= \left[\phi_1, i\sigma_2 \phi_2^*\right],\quad \phi_i= \left( \begin{array}{c} \phi_i^0 \\ \phi_i^- \end{array}\right),\quad i=1,2. 
\end{equation}
 The most general vev of 
$\Phi$ is given by
\begin{equation}\label{Phivev}
\langle\Phi\rangle=v\, \text{diag} (\cos\beta,-\sin\beta e^{-ia})
\end{equation}
 The amount of spontaneous CP violation is measured by the small parameter $s_a t_{2\beta}$, with $s_a t_{2\beta}\lesssim 2m_b/m_t$~\cite{Senjanovic:2014pva}.

 Under parity, consistent with \eqref{ellunderP}, one has as 
\begin{equation}\label{parity}
 \Delta_L\leftrightarrow \Delta_R,\quad \Phi\rightarrow\Phi^{\dagger}
\end{equation}
so that  the Yukawa couplings of the bi-doublet are Hermitian and the left and right triplet Yukawas are the same.

The CP violating parameter $s_a t_{2\beta}$  measures the difference between  right and left-handed quark mixing matrix and thus controls the weak contribution to the strong CP violating parameter $\bar \theta$. It turns out that $s_a t_{2\beta}$ is practically vanishing~\cite{Maiezza:2014ala} (see however \cite{Kuchimanchi:2014ota}) in order to keep $\bar \theta$ acceptably small. The point is that with the spontaneously broken parity the strong CP parameter $\bar \theta$ is finite and calculable in perturbation theory~\cite{Beg:1978mt}. The crucial ingredient is the RH analog $V_R^q$ of the CKM matrix $V_L^q$ which is rather sensitive to $s_a t_{2 \beta}$. In recent years we had managed~\cite{Senjanovic:2014pva} to solve the long-standing problem of computing analytically $V_R^q$  which has troubled the MLRSM for some forty years. It turns that $V_R^q$ takes  a simple approximate  form
\begin{equation}
\label{eq:VR}
(V_R^q)_{ij} \simeq (V_L^q)_{ij} - i s_a t_{2 \beta}  \frac{(V_L^q)_{ik} ( V_L^{q \dagger}m_uV_L^q)_{kj} }{m_{d_k}+m_{d_j}}  
+O(\epsilon^2) 
\end{equation}
 It can be shown that the left and right mixing angles are almost the same, and right-handed phases depend only on $V_L$ and  
$s_a t_{2 \beta}$. Thus by measuring $V_R^q$ one can predict the amount of parity violation in the gauge interactions of quarks.  In particular the near equality of LH and RH quark mixing angles justifies the experimental limits on $W_R$ mass~\cite{Aaboud:2019wfg}. The knowledge of $V_R^q$ leads furthermore to precise predictions for low energy processes, see e.g.~\cite{Bertolini:2019out}.

The parity conserving limit $s_a t_{2 \beta} = 0$, motivated by the smallness of strong CP violation, is particularly clean since then one has the exact equality of the LH and RH quark mixing matrices. There are various cross-checks of the theory since many other interactions depend on $\epsilon$, in particular the ones of the heavy doublet residing in a bi-doublet. The parity conserving limit we are discussing here is well defined both theoretically and experimentally and may be of great phenomenological importance.

The SM symmetry breaking through $\langle \Phi \rangle$ induces a tiny vev $v_L$ of 
 the left-handed triplet $\Delta_L$ with  a hierarchy of $SU(2)_L$ breaking
$v_L\propto v^2 /v_R$~\cite{MohSenj81}. The naturally small $v_L$ is self-protected~\cite{MohSenj81} (for a recent discussion, see \cite{Maiezza:2016ybz}) and is a direct source of neutrino mass, coined type II seesaw~\cite{typeII,MohSenj81}.

\section{Seesaw and how to probe it}
\label{section:crux}

  Let us first diagonalise the charged lepton mass matrix  
  $M_e = E_L m_e E_R^\dagger$, by rotating the LH and RH doublets 
$\ell_L  \rightarrow E_L \ell_L, \,\,\, \ell_R \rightarrow E_R \ell_R$. The PMNS matrix $V_L$ and its right-handed counterpart  $V_R$ are then the unitary transformations that diagonalize $M_\nu$ and $M_N$, respectively. 

In general one has $E_L \neq E_R$, and the unitary matrix  
$U_e = E_R^{\dagger} E_L$ measures the amount of parity breaking. 
We have discussed this in detail in~\cite{Senjanovic:2018xtu}; here we focus on the situation of unbroken (or very weakly broken) parity with $U_e=  \mathbb{I}  \,\,(\text{up to signs)}$ and 
\begin{equation}\label{MDhermitian}
M_D=M_D^{\dagger}.
\end{equation}

\subsection{Unbroken parity and the seesaw} \label{unbrokenP}

 In this case $M_D$ can be found analytically as a function of light and heavy neutrino mass matrices. In our previous work~\cite{Senjanovic:2018xtu} the computation of $M_D$ may appear somewhat mysterious and dependant on the mathematical approach taken. We clear now this issue in detail and show how $M_D$ gets determined independently of the approach taken.

     The starting point is the $(\nu_{L}, N_{L})$ mass matrix~\cite{Senjanovic:2018xtu}
\begin{equation}\label{nuNmassmatrix}
\left( \begin{array}{c c}\dfrac{v_L}{v_R} M_N^*& M_D^T \\[10pt] M_D & M_N \end{array}\right),  
\end{equation}
 a mixture of both type II and type I seesaw matrices. It can  be readily block-diagonalized in the seesaw assumption $M_N\gg M_D$ by the approximate unitary rotation (to the leading order in $M_D/M_N$)
\begin{equation}
 \left( \begin{array}{c} \nu \\ N \end{array}\right)_L \rightarrow 
 \left( \begin{array}{c c}  1 &  \Theta^{\dagger}   \\  -\Theta & 1   \end{array}\right)   \left( \begin{array}{c} \nu \\ N \end{array}\right)_L
\end{equation}
with
\begin{equation}\label{theta}
\Theta=\frac{1}{M_N}M_D
\end{equation}
The physical meaning of $\Theta$ is clear - it measures the mixing of light and heavy neutrinos and thus allows to predict 
decays $N \to W^{\pm} \ell^{\mp}$ (or $W \to  N \ell$ if $W$ is heavier than the $N$). Our aim is to compute it as a function of $M_\nu$ and $M_N$, which is equivalent to computing $M_D$. More about it later.

This in turn leads to the neutrino mass matrix to the leading order in $M_D/M_N$
  \begin{equation}\label{seesaw}
M_{\nu}=\frac{v_L}{v_R} M_N^* -M_D^T\frac{1}{M_N}M_D
\end{equation}

  This is the celebrated seesaw formula, but the question is how to interpret it? If one had a fundamental theory of $M_D$, say relateting it to the charged lepton or quark mass matrices, one could predict $M_\nu$, once $M_N$ was known. 
  However, it is $M_\nu$ that is being measured today, and together with $M_N$, should be used as an input in order to probe its seesaw origin. The rest of this work is devoted precisely to the task of determining $M_D$ from the above formula. Before we plunge into it, a few words about probing the RH neutrino mass matrix $M_N$.

   \subsection{Determining light and heavy neutrino mass matrices}\label{MnuMN}

  These LH and RH neutrino mass matrices can be diagonalized by the unitary rotations $V_L$ and $V_R$, respectively. One writes
$M_{\nu}=V_L^*m_{\nu}V_L^{\dagger}$,
where $m_\nu$ stands for diagonal neutrino masses and $V_L$ is the standard PMNS mixing matrix. This amounts to a rotation $\nu_{L} \rightarrow V_L\nu_L$ when going from the weak to the mass basis. 

Similarly, 
$M_N=V_Rm_NV_R^T$,
where $m_N$ stands for diagonal matrix of heavy neutrino masses. This means $N_{L} \rightarrow V_R^* N_L$ (or $\nu_{R} \rightarrow V_R \nu_R$ in analogy with the LH neutrinos). 
In what follows we will be focusing on disentangling the seesaw, i.e. finding $M_D$ from $M_{\nu}$ and $M_N$. The light neutrino mass matrix $M_\nu$ is slowly but surely being determined from neutrino oscillations, and together with other low energy processes such as neutrinoless double beta decay, the electron end-point energy experiments such as KATRIN, the JUNO and Dune experiments and others, one has a realistic hope of knowing both $m_\nu$ and $V_L$ in a foreseeable future. 

The heavy neutrino mass matrix $M_N$ is  to be extracted from hadron colliders such as the LHC, through the production of $W_R$ and $N$ (the KS process)  and the production of new scalars of the theory, especially the double charged ones and their decays which depend on $m_N$ and $V_R$.
There is still a possibility - albeit less appealing - that these particles are too light to be seen at colliders. This would be true in particular if the lightest $N$ were to be the warm dark matter as in the Dodelson-Widrow scenario~\cite{Dodelson:1993je}. It is noteworthy that in this case  $M_N$ gets fixed completely, with practically zero mixings and masses in the keV-GeV range~\cite{Nemevsek:2012cd}. Moreover, $W_R$ is either too heavy to be seen at the LHC or it must live in a tiny window $M_{W_R} \simeq 5\,$ TeV. In this case the $W \to N \ell$ decays become potentially observable, especially when $N$ is precisely the lightest RH neutrino, the DM candidate.

   
 \subsection{Unbroken parity: untangling the seesaw}\label{untangle}  

Here we show how the seesaw gets disentangled in the MLRSM.  All that the reader  needs is the above seesaw formula of neutrino masses in \eqref{seesaw} and  property \eqref{MDhermitian}; this
 suffices to solve for $M_D$ and in turn predict physical decay rates. The argument goes as follows.

Firstly, for simplicity and transparency we introduce the Hermitian matrix $H$ defined through
\begin{equation}\label{Hdef}
M_D = \sqrt{M_N}  \,H \sqrt{M_N^*}
\end{equation}
Next, using this definition in \eqref{seesaw},  taking the complex conjugate  and dividing $\sqrt{M_N}$ on both sides, one readily obtains an symmetric matrix equation for $H$%
\begin{equation}\label{HHT}
H H^T = S,
\end{equation}
where the  symmetric matrix $S$ is given by (recall that Majorana mass matrices as well as their square roots are symmetric)
\begin{equation}\label{S}
S = \frac{v_{L}^*}{v_R}-\frac{1}{\sqrt{M_N}}M_{\nu}^*\frac{1}{\sqrt{M_N}}.
\end{equation}

The physical meaning of $H$ is clear: it is simply a $\nu-N$ mixing matrix $\Theta$ made Hermitian. Once $H$ is known, it is straightforward to find $M_D$ and $\Theta$.

A comment is noteworthy at this point. Due to parity symmetry, there are a series of constraints to take into account. Since $\text{Im}\,\text{Tr}\,\left(HH^T\right)^n = 0$ for any $n$ and Hermitian $H$, 
one has the following conditions
 \begin{equation}\label{conditions}
  \text{Im}\,\text{Tr}\left[ \frac{v_{L}^*}{v_R}-\frac{1}{M_N}  M_{\nu}^*\right]^n=0, \quad n=1,2,3.
\end{equation}
These constraints will play an important role in simplifying our results. Their mathematical meaning is clear: the coefficients of the characteristic polynomial of the matrix $S$ are real, implying real or pairs of complex conjugate eigenvalues. 

In all of this, the input  physical matrices $M_\nu$ and $M_N$ are measured at low energies and through KS process at hadron colliders, respectively. Just as you input the charged fermion mass in the SM in order to get the corresponding Yukawa coupling and the associated Higgs decays, in the Majorana picture of neutrinos one needs both light and heavy neutrino masses, independently. But the essence is exactly the same, as long you can predict $Y_D = M_D/v$. This is what aim for in what follows.

 \vspace{0.2cm} 

We now offer various ways of solving for $H$.

\vspace{0.2cm} 

{\bf Direct approach.} 
 The $n^2$ elements of $H$ can be found directly from \eqref{HHT}. It is enough to take into account the $n$ conditions \eqref{conditions} on $S$ to have a solvable system of $n^2$ second order equations for $n^2$ variables. But before rushing into the straightforward calculation, we show how the problem can be significantly simplified.
 From \eqref{HHT}, by using the identity $(H H^T) H = H (H^T H)$ and the hermicity condition $H ^T = H^*$ one gets
 \begin{equation}\label{HandS}
 S H =  H S^* 
\end{equation}
The above linear equation are easily solvable for  $n^2-n$ elements of $H$.  The rest of the $n$ elements of $H$ can then be found by using  \eqref{HHT}, this time reduced to a system of $n$ (rather than $n^2$) equations of second degree. In general, the number of discreet solutions equals  $2^n$. This procedure determines  $H$,  from where it follows $M_D$ in \eqref{Hdef}  and  the $\nu-N$ mixing $\Theta$ as defined in \eqref{theta}. 

    Though in general there is no analytic expression for $H$, it is instructive to illustrate the situation for the simplifying case of two generations where one gets
    \begin{equation}
  H_{2\times2}=  \sqrt{SS^*}\frac{1}{\sqrt{S^*}}.
  \end{equation}
   Despite its non-manifestly hermiticity, it can be shown by inspection that this matrix  is indeed Hermitian
 for only four combination of the square roots and thus serves to untangle the seesaw.

  \vspace{0.2cm} 

  {\bf Jordan decomposition.} There is nothing wrong with working with the equations that determine $H$ directly, after all it is a well defined program of solving quadratic and linear matrix equations. However, having a simple analytic expression gives both more insight and  eases calculational pain, and it is achieved only by giving up the usual program of diagonalising a matrix (in this case the symmetric $S$) by an unitary transformation. The crucial step is to decompose the symmetric matrix \eqref{S} as in~\cite{Senjanovic:2016vxw}
  \begin{align}\label{Omatrix}
S= O\, s \,O^T
  \end{align}
where $O$ is a complex orthogonal matrix and $s$ is known as the symmetric normal form \cite{Gantmacher}. 
A comment is noteworthy here. It is customary to use a unitary matrix instead of the orthogonal one, since then a symmetric matrix is guaranteed to be diagonalised. The unitary matrix approach, though, does not work in this case as we discuss below in detail. It is important to keep in mind, however, that the symmetric normal form $s$ will in  general not be diagonal.

Since $H$ is in general complex, solving the above equation turns out non-trivial. But notice that due to the hermicity of $H$, equation \eqref{HHT} can be written as $H H^* = S$, and $H$ would have to be real to allow taking the square root, which would give
 $H = O \sqrt s\, O^T$  in the notation of  \eqref{Omatrix}. In~\cite{Senjanovic:2016vxw} we have managed to show that for $H$ complex this generalises to a rather simple expression
\begin{equation}\label{detH}
H=O\sqrt{s}EO^{\dagger} 
\end{equation}
In general, $E$ would not be easy to determine, but since $H=H^{\dagger}$, one can deduce the following conditions 
 \begin{equation}\label{Hfound}
\sqrt{s}E=E\sqrt{s^{*}}, \quad \, E^T=E^*=E^{-1}.
\end{equation}
It is an easy exercise to show that the  equations for $E$ can be simplified to
 \begin{equation}\label{Hfound-2}
{s}E=E {s^{*}}, \quad \, E^T=E^*=E^{-1},
\end{equation}
which will be used hereafter.

 By computing $E$,  one can achieve the task of disentangling the seesaw by determining  $M_D$   as     
\begin{equation}\label{MDdis}
M_D=\sqrt{M_N}\,O\,\sqrt{s}\,E\,O^{\dagger}\sqrt{M_N^*}
\end{equation}
Since $O$, $s$ and $E$ all follow from $M_\nu$ and $M_N$,  this manifestly shows how in the parity conserving case $M_D$ can be determined from the solely knowledge of the light and heavy neutrino masses and mixings.

\vspace{0.2cm}

{\bf Unitary diagonalisation.}
  The reader may ask what happens if one sticks to the traditional approach and diagonalise $S$ by means of an unitary transformation, say   
 \begin{equation}\label{OH}
S = V d\, V^T
\end{equation}
In this case, analogously to \eqref{detH}, we can write
\begin{equation}\label{detHV}
H = V \sqrt{d\, V^TV}  E_V V^\dagger
\end{equation} 
where $d$ stands for the diagonal real matrix, $V V^\dagger =1$ and $E_V$ is an Hermitian matrix  needed to ensure an Hermitian $H$. Unlike the previous approach  which  makes use of the Jordan decomposition and which led  to the simple form of \eqref{detH}, here  the factor $V^TV$ in \eqref{detHV} remains entirely and does not reduce to the unit matrix as before.
The rest of the procedure follow straightforward. The hermiticity condition for  $H$ provides the following conditions on the matrix $E_V$
\begin{equation}
d\, V^TV E_V=E_V  V^{\dagger}V^*d 
\end{equation} 
and the less transparent
\begin{equation}
E^T_V=E^*_V=V^TV E^{-1}_V V^{\dagger}V^*
\end{equation}
in complete analogy with \eqref{Hfound-2}. 

At this point, we can clearly see the advantage of the Jordan decomposition over unitary diagonalization. Even with $d$ guaranteed to be diagonal, the non simplification of the factor $V^TV$, due to the presence of complex phases in $V$, makes this approach not suitable for our purpose. In what follows we continue with the Jordan decomposition and show how it can be used to find  all possible $H$.

\subsection{Jordan decomposition: explicit expressions}\label{explicit}

Expression \eqref{MDdis} is valid for any $M_\nu$ and $M_N$, i.e., any normal form $s$, which is either diagonal or not. We will treat these two cases separately, and show how the Jordan decomposition procedure fixes the possibilities for $s$ and the corresponding $E$ matrices, which then determine $H$ and in turn $M_D$. This will nicely illustrate the power of the Jordan procedure.
\vspace{0.2cm} 

(i) {\bf Diagonal normal form}. Let us start first with the simpler situation of diagonal $s$. In this case the constraints \eqref{conditions} allow, in the $3\times3$  case, only two distinct possibilities
\begin{equation}\label{s-diagonal}
  s_{I}=\text{diag}(s_1,s_0,s_2) ,\quad   s_{II}=\text{diag}(s,s_0,s^*) 
    \end{equation}
with $s_{0,1,2}$ being real numbers.  From \eqref{Hfound-2} and \eqref{s-diagonal}
the matrix $E$ are  found to be 
  \begin{equation}
E_{I}=  \left(   \begin{array}{ccc}
 1& 0&0 \\ 
  0& 1 & 0 \\ 
0 & 0 & 1
\end{array}   \right), \qquad  E_{II}=  \left(   \begin{array}{ccc}
 0& 0&1 \\ 
  0& 1 & 0 \\ 
1 & 0 &  0
\end{array}   \right)
    \end{equation}
corresponding to the values of $s_I$ and $s_{II}$, respectively.
 The $H$ matrix is then given explicitly by
\begin{equation}
  H_{I}= O \left(   \begin{array}{ccc}
  \sqrt{s_1}&0 &0 \\ 
0 &  \sqrt{s_0} & 0 \\ 
0 & 0 &  \sqrt{s_2}
\end{array}   \right) O^{\dagger},
 \end{equation}
and 
\begin{equation}   
H_{II}=    O \left(   \begin{array}{ccc}
 0 &0 &\sqrt{s} \\ 
0 & \sqrt{s_0} & 0 \\ 
\sqrt{s}^* & 0 & 0
\end{array}   \right) O^{\dagger},
    \end{equation}
    corresponding to $s_I$ and $s_{II}$, respectively.
There are eight possibilities in the first case, four in the second.

Equation \eqref{s-diagonal} can be generalized to any number of generations $n$: for $n$ even, for every eigenvalue $z$, there is also an eigenvalue $z^*$.
For $n$ odd there is on top one real eigenvalue. The matrix $E$ in this case has a 1 in the diagonal  for each corresponding 
real eigenvalues and two 1's  symmetrically opposed in the anti-diagonal (anti-diagonal being defined from the lower left corner to the upper right corner) for each corresponding complex eigenvalue and its conjugate.

\vspace{0.2 cm}

(ii) {\bf Non-diagonal normal form}. The next and final case is when $s$ non-diagonal. Again, for the $3\times3$ case Jordan tells us that there are only two alternatives which exhaust the possible symmetric Jordan normal forms, namely a $2 \times 2$ block combined with a single entry element, and a single $3 \times 3$ block. These symmetric blocks can be obtained from the canonical normal blocks by mean of a simple transformation, see~\cite{Gantmacher}
\begin{equation}\label{nondiagonal-s-block}
s_{block}=\frac{1}{2}(\mathbb{I} - i \mathbb{I_A}) \left(s_i \mathbb{I} + \mathbb{E} \right)(\mathbb{I}+ i \mathbb{I_A})
\end{equation}
where the matrices $\mathbb{I}$ and $\mathbb{I_A}$ are  diagonal and anti-diagonal with unit entries, respectively, while $\mathbb{E}$ is a matrix in which the only non-zero elements, equal to 1, lie immediately above the main diagonal (normally called the superdiagonal of a square matrix). The second factor in \eqref{nondiagonal-s-block} represents a general block of the canonical normal form.

With this in mind, we can now write all possible non-diagonal symmetric normal forms for  $3\times3$ matrices 
    \begin{equation}\label{nondiagonal-s}
s_{III}=  \left(   \begin{array}{ccc}
 s_1+\dfrac{i}{2}& \dfrac{1}{2}&0 \\ [8pt]
  \dfrac{1}{2}& s_1 -\dfrac{i}{2}& 0 \\ [8pt]
0 & 0 & s_2
\end{array}   \right)
    \end{equation}
and
 \begin{equation}\label{nondiagonal-s}
 s_{IV}=  \left(   \begin{array}{ccc}
 s_0& \dfrac{1+i}{2}&0 \\[8pt] 
  \dfrac{1+i}{2}& s_0 & \dfrac{1-i}{2} \\[8pt] 
0 &\dfrac{1-i}{2} &  s_0\\[5pt]
\end{array}   \right).
    \end{equation}
    Since $\text{Tr}\, s_{III}^n=2s_1^n+s_2^n$ and $\text{Tr} \,s_{IV}^n=3s_0^n$, the reality conditions \eqref{conditions} imply real $s_0,s_1$ and $s_2$.
Proceeding analogously as in previous case, the matrix $E$ can be found  from \eqref{Hfound-2} and \eqref{nondiagonal-s} and are given by
  \begin{equation}\label{Ematrix-III-IV}
E_{III}=  \left(   \begin{array}{ccc}
 0& 1&0 \\ 
  1& 0 & 0 \\ 
0 & 0 & 1
\end{array}   \right), \qquad  E_{IV}=  \left(   \begin{array}{ccc}
 0& 0&1 \\ 
  0& 1 & 0 \\ 
1 & 0 &  0
\end{array}   \right)
    \end{equation}
for the respective values of $s_{III}$ and $s_{IV}$. To find $H$, it remains to compute the square root of $s_{III}$ and $s_{IV}$. For this, it is enough to focus on the Jordan blocks only. The square root of an arbitary block can be found directly from \eqref{nondiagonal-s-block}
\begin{equation}\label{root-nondiagonal-s}
\sqrt{s_{block}}=\frac{1}{2}(\mathbb{I} - i \mathbb{I_A}) \left(\sqrt{ s_i}\,\mathbb{I}+\frac{\mathbb{E}}{2s_i^{\frac{1}{2}}} -\frac{\mathbb{E}^2}{8s_i^{\frac32}} +\cdots\right)(\mathbb{I}+ i \mathbb{I_A})
\end{equation}
The above series in parenthesis  breaks off after the $n$th term for an $n\times n$ block and it follows from the expansion of the square root of the canonical Jordan block as a power series in 
$\mathbb{E}$.

For the $3\times3$  case, the  $H$ matrices are then found to be 
\begin{equation}\label{HIII}
H_{III}= O\left(\!\! \begin{array}{ccc}
  \dfrac{1}{4\sqrt{s_1}} & \sqrt{s_1}\!+\!\dfrac{i}{4\sqrt{s_1}} &0 \\ [10pt]
   \sqrt{s_1} \!-\!\dfrac{i}{4\sqrt{s_1}} & \dfrac{1}{4 \sqrt{s_1}}& 0 \\ [10pt]
0 & 0 & \sqrt{s_2}
\end{array}   \right)O^{\dagger}   \end{equation}
 and
     \begin{equation}\label{HIV}
H_{IV}=      O \sqrt{s_0}\left(   \begin{array}{ccc}
 -\dfrac{1}{16s_0^2}  & \dfrac{1+i}{4 s_0}   &1\!-\!\dfrac{i}{16  s_0^2 } \\[10pt] 
 \dfrac{1-i}{4s_0}            & 1      & \dfrac{1+i}{4 s_0 } \\[10pt] 
  1\!+\!\dfrac{i}{16  s_0^2 }                              & \dfrac{1-i}{4 s_0}      &  -\dfrac{1}{16 s_0^2}\\[5pt]
\end{array}   \right)O^{\dagger}  
 \end{equation}
 which correspond to the matrices $s_{III}$ and $s_{IV}$, respectively.

Notice that due to the particular form of the matrices $s_{III}$ and $s_{IV}$, the elements of the above matrix $O$ have to be large in order to amount for the  small ratio of left and right-handed neutrinos masses, as required by \eqref{S}. For the diagonal case this is not true - the orthogonal matrix becomes  the unit matrix when the left and right-handed mixing angles are equal~\cite{Senjanovic:2018xtu}. 
  
It the same way as in the diagonal case,  it is also possible to compute $H$ for  higher number of generations following the above procedure. The number of possible non-diagonal symmetric blocks and their combination will grow  and can  easily be calculated for each particular dimension.  The general formulas \eqref{nondiagonal-s-block} and \eqref{root-nondiagonal-s} are valid in any dimension, and the task is greatly simplified by  conditions \eqref{conditions}, which tell us that $s_i$ are either real or come in complex conjugate pairs.

 A final comment. We have stayed away here from singular points in which the  square root is ill-defined. 
\vspace{0.2cm}

   In summary, it is evident that the the Jordan decomposition of the $S$ matrix allow us to get all the analytical solutions for $M_D$ in a simple and compact matter.

\section{Summary and outlook}
\label{section:outlook}

The origin and nature of neutrino mass is arguably one of the central issues in the quest for the theory beyond the Standard Model. Over the years, the seesaw mechanism has emerged as the main scenario behind the smallness of neutrino mass, but it suffers from two serious setbacks. Firstly, the SM seesaw cannot be disentangled, and secondly, the heavy RH neutrinos cannot be produced at  hadron colliders, such as the LHC, unless the Dirac mass terms are
 incomparably larger than their natural tiny values. 
 
 Both of these problems disappear in the context of the LR symmetric theory which attributes the breakdown of parity in weak interactions to its spontaneous origin. For this is sufficient to have RH neutrinos  produced at hadron colliders through the KS process which allows us to probe the masses and mixings of the heavy RH neutrinos. This is doubly checked through the production of double charged scalars in the triplet Higgs multiplets. 
 Moreover, in the MLRSM the Dirac mass terms are determined unambiguously, providing a testable Higgs theory of neutrino mass.  In the case of charge conjugation this  is quite straightforward since it keeps the Dirac mass matrix symmetric which immediately provides a solution for $M_D$ as a function of $M_\nu$ and $M_N$, the light and heavy neutrino mass matrices, respectively.
 
 The case of parity turned out to be much more difficult since in general $M_D$ is neither symmetric nor Hermitian. Nonetheless, we have managed to provide a way of dealing with this and we have also found a  solution for $M_D$ in the limit of unbroken parity in the Dirac Yukawa sector.
 In this work, we have gone one step beyond. We have completed this program by providing all possible solutions for the Dirac Yukawa mass matrix (or equivalently the Dirac Yukawa couplings) as a function of $M_\nu$ and $M_N$. The seesaw mechanism, in conclusion, gets completely disentangled in the MLRSM.
 
The reader should justifiably raise the question of the scale though: why should the MLRSM be accessible at the LHC or next hadron collider energies? The answer lies as we argued in a deep connection with a neutrinoless double beta decay whose observation may signal the contribution of new physics if neutrino mass is not sufficient to do the job. In this case $W_R$ could not be too heavy~\cite{Nemevsek:2011aa} making the case for its manifestation at the LHC, for a recent in-depth study see~\cite{Nemevsek:2018bbt}. Moreover the KS process also provides a direct lepton flavor violation and is thus connected to low energy analogous processes~\cite{Tello:2010am}

%
%
%

 A few concluding remarks regarding what it means to have a theory of neutrino mass. One often adds additional discrete or continuous symmetries to a gauge symmetry in question, with the purpose of determining $M_D$. 
   In the LRSM this is not needed since the theory does the job anyway due to its internal structure. 
   In other cases, one chooses the parameter space of the theory in advance, but that requires abandoning minimality and losing original predictions. 
   
  Recall that good theories typically have a contrived parameter space; e.g. in the SM the heavy top quark requires a tiny mixing between the first and the third generation. What we advocate  here is to keep the minimality and predictivity to the bitter end and let the experiment have the final word.

 In summary, the great success of the SM in accounting for particle masses is based on the maximal breaking of parity symmetry. By restoring parity one cures its failure to account for the non-vanishing neutrino mass in a self-contained and predictive manner. 
 

\vspace{0.6cm}

\subsection*{Acknowledgments}

  We wish to acknowledge the collaboration with Miha Nemev\v{s}ek on the original study of neutrino mass with parity.  GS wishes to thank the Fermilab theory group, especially Stephen Parke for making him feel at home, and the TD Lee institute, for their warm hospitality during various stages of this work. We are grateful to Gia Dvali, Alessio Maiezza and Juan Carlos Vasquez for fruitful discussions and comments.

\end{document}